# Fourier optics for polymeric substrates and coating textures analysis.

**Amelia Sparavigna**
**Physics Department, Politecnico di Torino**
**C.so Duca degli Abruzzi 24**
**Torino, 10129 Italy**

**Rory A. Wolf**
**Enercon Industries Corp.**
**Menomonee Falls, Wisconsin, USA**

**Abstract:** Several devices for substrate texture detection based on diffractive optics, for paper, textiles and non-wovens have been proposed in the past for direct inspection during the production processes. In spite of the presence of devices totally based on image processing, the use of diffractive optics cannot be considered surpassed for many reasons. Compared with image processing procedures, it is less sensitive to vibrations and does not suffer from the presence of ambient light. Based on transmitted light, it can give information on changes in refractive indexes, thickness variation and surface conditions. We study the use of optical Fourier spectrum to identify textures of polymer films. As the power spectrum reveals, the texture is seldom homogeneous. Here we report investigation on several substrates and on thin ink coatings on substrate. Role of bulk and surface conditions is analysed.

**Keywords:** Fourier optics, texture detection, printing analysis.

## 1. INTRODUCTION

For a direct inspection during the production processes of paper, textiles and non-wovens, several systems based on the image processing are now available. Devices based on the Fourier optics represent an interesting alternative for inspection. In fact the common trend prefers image-processing systems, but a combined use with Fourier optics could be convenient in some circumstances. Compared with image-processing procedures, Fourier optics analysis is less sensitive to vibrations and does not suffer from the presence of ambient light. It works in transmitted light and then provides information on changes in refractive indexes, thickness variation and surface conditions.

Optical Fourier transform (OFT) of a transparent object can be simply obtained with a laser beam and a lens. The object, for instance a transparency on which an image is reproduced, is placed in front of a lens and lightened with a parallel beam of coherent light: on the rear side of the lens, at the focal distance, the Fourier power spectrum of the image is formed [1-4]. The spectrum can be detected with a CCD camera. Fourier optical analysis is fast and quite suitable for both off-line and on-line analysis. It is a good method for monitoring orientation distribution of fibres and checking regularity in substrate textures and for detection of defects. In the case of fabrics for instance, we can identify the defect according to its deviation from the regular weaving of yarns. The same is true for papers and non-wovens [5-7].

Lendaris and Stanley introduced the use of Fourier optics, for the first time in 1970. to analyse photographic imagery [8]. In the same year, the use of OFT was proposed to determine the structure of paper: further developments allowed determining the orientation of fibers in thin paper samples. Many other applications were then proposed with more or less good results. H.L. Kasdan was among the first researchers to study industrial OFT systems for paper quality evaluation and for detection of defects in textile fabrics [9]. For fabrics, OFT power spectrum has a pattern characterised by a set of peaks. Because the spectrum of a fabric with a defect is different from that of the good fabric (peaks changes in intensity, shape and position), it is possible to develop an inspection system, directly on the loom. Recently, investigations on the structure of non-woven fabrics have been reported in the International Nonwovens Journal, 2001, dealing with various methods to characterise nonwovens, measuring the orientation distribution function and basis weight [10]. The observations of researchers show that OFT processing is unquestionable especially when the sample weight/unit area was high.

Optical Fourier spectrum can be used to analyse textures of polymer films. Spectra reveal that the structure of polymer films is seldom homogeneous and this can be attributed to refractive indexes and surface morphology variations, due to the material and to the overall processes at which polymer films are subjected.

## 2. DIFFRACTIVE OPTICS FOR FAULT DETECTION

OFT find the best application in those devices where the check of regularity in an image texture is required. In the case of fabrics for instance, because as previously told, defects are deviations from the regular weaving of yarns. Harvey L. Kasdan studied how to find the most common defects in cotton fabrics: missing yarn, double pick and density change in the weaving process of the web. A missing yarn happens when a yarn breaks in the weaving mechanism of the web. A consequent deviation from the regular geometry of the fabric is produced, because the distance between yarns is changed. Figure 1, on the left, shows that the lack of a yarn gives a quite visible defect. Another defect shown in the same figure, on the right, is produced when the spacing among yarns is reduced due to a density change in the weaving of the web. The double pick happens when two picks are in the same weaving position of the fabric and it is rather difficult to detect.

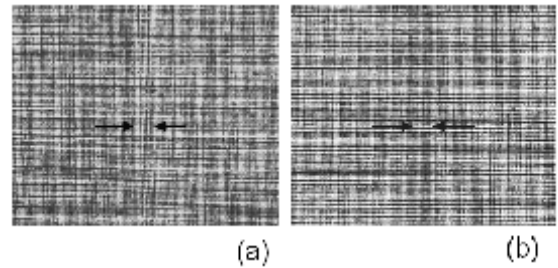

**Figure 1**: Defects in a fabric. On the left, the defect produced by a missing yarn, on the right the result of a density change in the weaving process. The image size corresponds to 3 cm.

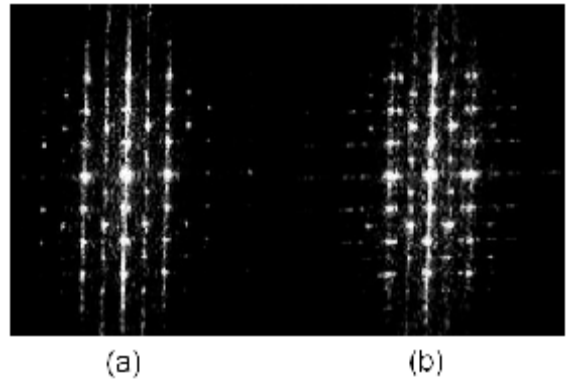

**Figure 2**: Power spectrum of the fabric without (a) and with a double-pick defect (b).

The work of Kasdan shows that it is possible to detect defects with an optical Fourier transform (OFT). The power spectrum of a fabric with a defect is different from that of the good fabric: peaks changes in intensity, shape and position. In Fig.2, we see that the double pick, as observed by Kasdan, gives a splitting of some peaks in the Fourier transform. To have an automatic evaluation of changes in the peaks, Kasdan used an array of photo-detectors.

Several other approaches for applying OFT to fabric inspections have the proposed [12-14]. Let us conclude the section, telling that a direct on-loom inspection of fabrics is hard to obtain, both with image-processing systems and OFT systems.

## 3. EXPERIMENTAL SET-UP AND RESULTS

An experimental set-up for OFT detection is shown in Fig.3. The system is composed with a diode laser in front of a beam expander, two mirrors and the optics for the Fourier transform. The power spectrum is formed on a CCD camera detector. The intensity of the beam can be adjusted with a polarizer. The light source is a laser-diode with wavelength at 670 nm. The beam expander gives a light beam with a diameter of 30 mm. A stop with an adjustable diameter is placed after the beam expander, to suitably change the diameter of the light spot. The Fourier optics is composed of two lenses and has a focal length of 170 mm. Optics 1 is the laser lens and 2 the beam expander. A filter (element 6) is placed between elements 5 and 7 of the Fourier optics, to remove ambient light. 3 and 4 are mirrors.

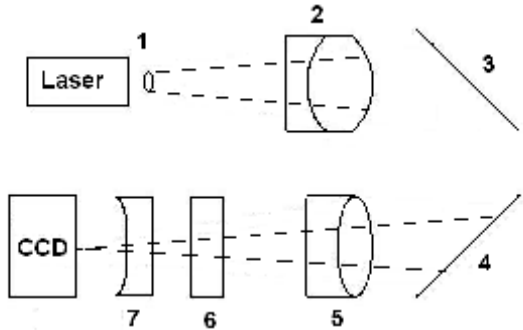

**Figure 3**: Optical set-up. 1 is the laser lens and 2 the beam expander. Filter 6 is placed between 5 and 7, to remove ambient light. 3 and 4 are mirrors.

On the CCD, it is created the power spectrum of an object placed between the two mirrors. It is necessary to observe that, with the recording by means of CCD camera and computer, we have not the true spectrum coming from the Fourier optics. Due to camera and frame-grabber shutters, used to capture and record the images, the recorded intensity is slightly different from that produced by the Fourier optics.

An object is located in the $(x, y)$-plane in front of the Fourier lens, at a distance $d_o$ and lightened by a normally incident plane wave of amplitude $A$. A function $f(x, y)$ can be used to represent the light transmission. $\lambda$ is the wavelength of the laser and $f_c$ the focal distance of the Fourier optics. The Fourier Transform produces an intensity distribution across the focal plane, given by:

$$I(x,y) = \frac{A^2}{\lambda^2 f_c^2} exp\left[ j\frac{k}{f_c}\left(1 - \frac{d_o}{f_c}\right)(u^2 + v^2)\right] \times$$

$$\left[ \iint f(x,y) exp\left[ -j\frac{2\pi}{\lambda f_c}(xu + yv)\right]dxdy\right]^2 \tag{1}$$

where $(u, v)$ are the coordinates in the focal plane. The wave-number $k$ is $2\pi / \lambda$. The frequencies in the power spectrum are given by the following expressions:

$$f_x = \frac{u}{\lambda f_c} \quad ; \quad f_y = \frac{v}{\lambda f_c} \tag{2}$$

The complex function $f(x, y)$ can simply represents a modulation in the intensity of light passing through the material; in the case of pure absorption, the function can be posed as real and used to model the fabric, as well as paper or nonwovens. For an investigation on polymer films, function $f(x, y)$ must consider absorption but also the presence of local variations in refractive indexes, the role of relieves on the surface of the sample, and changes in the thickness of the sample. These factors are described in the phase of a complex function $f(x, y)$. An important reason to use Fourier Optics is that it can give much more information on samples than a simple evaluation with a Fast Fourier processing of images.

The Fourier spectrum image is transferred from the CCD to a computer and visualised on a monitor. The image of the Fourier spectrum can be further elaborated. At authors' knowledge there is only one optical commercial system performing convolution between a good spectrum and a spectrum with defects [15].

## 4. POLYMERIC FILMS: ROLE OF BULK AND SURFACE.

Often the analysis of a polymer film with a microscope is not able to give information on its morphology. It is very difficult to observe the presence of textures and ever more to record images with a CCD camera. Moreover, many polymer films have an optical anisotropy due to the stretching of material during extrusion processes and microscopy in polarized light is required to see the textures. One could then erroneously conclude that polymeric films cannot be characterised with OFT, due to an apparently lack of textures in the substrate. In fact, some polymer films, placed in front of the Fourier optics, have the same behaviour of a glass, that is they do not modify the diffraction image of the light beam, but this is a special case. A quite different behaviour is more common, for instance, with polymer films of polypropylene and polyethylene.

Figure 4 shows the Fourier spectra of the laser source (a) and of three polymer films of biaxially oriented polypropylene (BOPP), with different thickness and coming from different producers. The halo of the OFT indicates the presence of an anisotropic structure in the bulk: the two directions displayed by OFT are not perpendicular. To record the OFT images, the samples are turned of 45 degrees with respect to CCD frame, to avoid any confusion with the laser peaks. Observing BOPP materials with polarized light microscope in crossed polarizer configurations, it is possible to see the textures. We used three samples (b),(c) and (d) (see Fig.4, lower part). Sample (b) is thick enough to show an enhancement of interference colours, to be appreciated in image recording [16]. By eye inspection, the surfaces of the three samples appear as a slightly undulating surface.

According to the previous discussion on the origin of the Fourier pattern, and in agreement with the observation by means of the polarized microscopy, we can conclude that the Fourier pattern is due to a modulation of the local optical density. This local density is related to a different local orientation of optical axis.

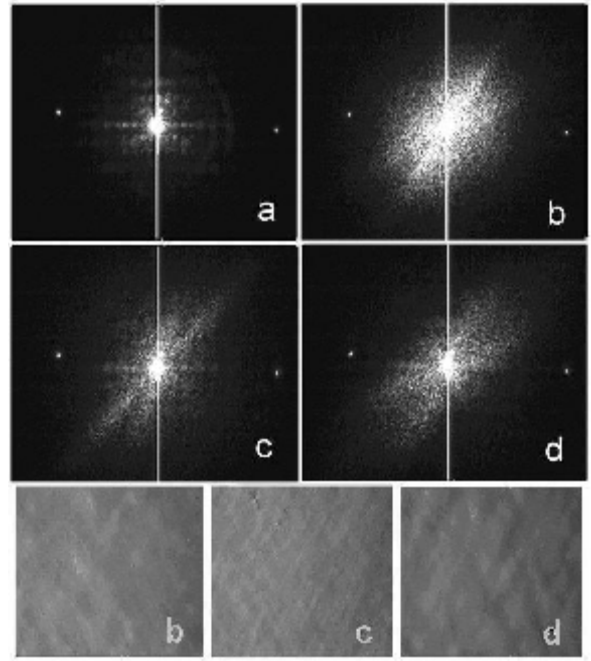

**Figure 4:** In the upper part, Fourier spectrum of the laser source (a) and of three polymeric films of BOPP, with different thickness (0.04 mm (b), 0.024 mm (c), 0.014 mm (d)). The halos indicate the presence of an anisotropic structure in the bulk. Images are recorder with a fixed intensity of the laser beam. In the lower part, the polarized light textures of the same three polymer films (image size, 1 mm). The anisotropic textures correspond to Fourier patterns.

We have also analyzed low-density polyethylene (LDPE) films. The OFT halo has an elliptic shape, more or less pronounced, depending on the origin of the sample: an example is shown in Fig. 5 on the left part. In the case of LDPE, a comparison with optical microscopy in polarized light is interesting too, since the microscope reveals an anisotropic texture, in agreement with the Fourier spectrum. In the same Figure 5, on the right, the polarized microscope texture is reproduced.

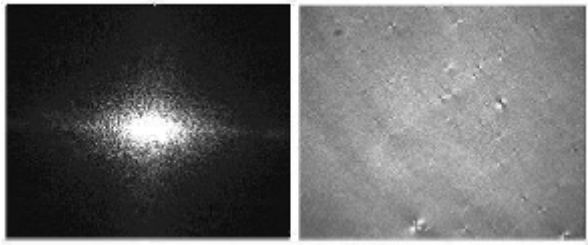

**Figure 5**: On the left, the Fourier spectrum of a polymer film LDPE compared with its microscope image on the right. The elliptic halo depends on the film texture, due to the overall production processes at which the film has been subjected.

For LDPE films, local changes of refractive indexes are possible as shown by the polarized light microscopy, but surfaces of these films are not flat. A combined action of refractive indexes and relieves on the surface can be ascribed as the origin of the OFT. The halo has an elliptic shape, due to an anisotropy in the optical axis induced by stretching and because the grooves on the surface have a preferred direction, depending on the production process.

The role of thickness and a non-smooth surface in forming the OFT is not negligible, as shown by the quite remarkable Fourier pattern for polypropylene with an orange-peel texture [16]. At link [16], we can see a pattern formed by dots diffused on all the image without a central peak. The image resembles a Speckle interference pattern of coherent light reflected by a rough surface. This is an utmost case, since the sample cannot be considered as a planar object. But, this stimulated us to perform another investigation, that on film coatings [17].

## 5. POLYMERIC FILMS: INK AND COATING.

Optical Fourier transform can describe the textures of coatings and inks. Let us consider for instance a polyester (PET) film. The polyester sample we observed has a thickness of 0.15 mm, a very flat surface and no textures are seen by microscope, except very few defects. Bulk and surface then give no contribution to Fourier pattern. Fig. 6(a) shows OFT of PET film: the film simply reduces the intensity of transmitted light by absorption, and the spectrum is the same of that observed for the light beam. Figure 6(b) shows the Fourier spectrum of the same polyester film, with a coating suitable for inkjet printers. The surface of this PET film for inkjet printer is covered with micro protuberances for anchoring purposes. These points are randomly distributed on the surface [16]. The coated film has a Fourier spectrum with a circular halo around the central peak. The halo is symmetric since the distribution of micro protuberances on the surface is uniform. From this example, we suppose that OFT can be used in some cases to determine and control, during processing, the texture of coatings.

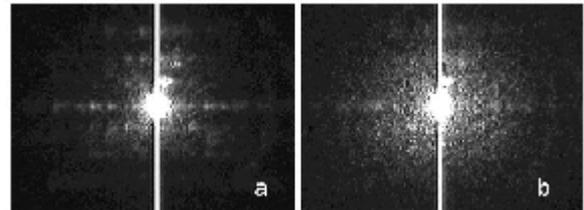

**Figure 6**: Fourier spectrum of the PET film (a) and of PET film coated for inkjet printers (b). The PET film does not alter the Fourier pattern of the source, just reduces the intensity. The coated film shows a halo due to the presence of micro protuberances evenly distributed on the surface [16]

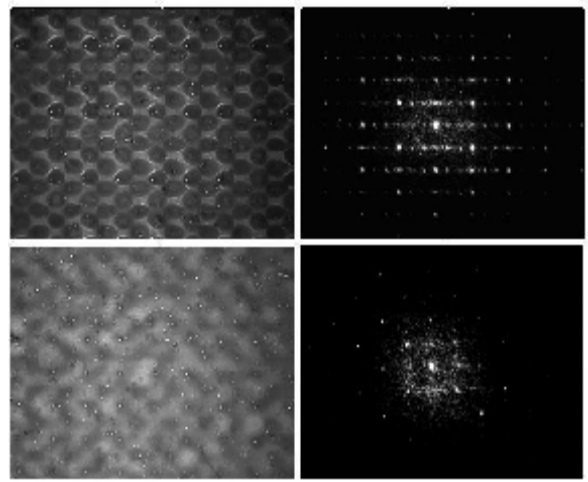

**Figure 7**: Rotogravure printed polymer film viewed with microscope and the OFT.

Investigations on printing processes of polymer films are possible too (see Fig.7), to determine the texture of inks and interaction of the ink coating with the surface. To print a polymer film, a rotogravure or a flexo technique, to cite among the others, can be used. Before printing a polymer film, the substrate can be subjected to a treatment with corona or plasma devices to increase the film wettability and to enhance the ink adhesion. In the case of rotogravure, the ink on the substrate assumes a periodic texture: the grooves on gravure rolls create a periodic dotted texture, clearly visible with the microscope observation. The Fourier spectra have patterns with regular peak positions, according to the periodicity of the ink dots [16]. Spectra are different according to the smoothness of textures. In this case, the Fourier optics works in the same manner as it works for fabrics. It can detect the periodicity of ink textures and defects.

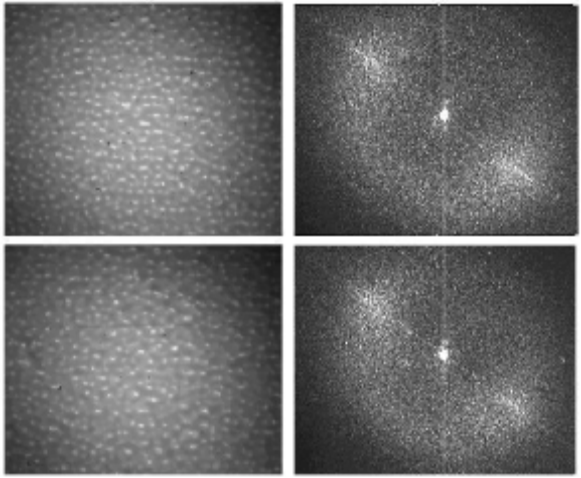

**Figure 8:** Two images of a flexo printed PET film observed with microscope (image size, 1 mm) and the corresponding Fourier spectra. A diverging beam is used to enhance in the microscope observation the orange peel structure of ink. In the upper part of the figure, the surface of the film was non-treated before printing. In the lower part, the surface was subjected to a plasma treatment to enhance the surface wettability. Fourier spectra are different.

In flexo printing there are no grooves on the printing roll. With a flexo roll, we printed a uniform image on a PET film and on a BOPP film (thickness 0.024 mm), with UV curable inks. The printed PET film observed with microscope is shown in Fig.8. A diverging light beam is used to enhance the orange peel structure of ink in microscope observation. On the right of Fig.8, the Fourier pattern is shown. In the upper part of the figure, the PET film under consideration was not subjected to pre-treatments before printing. In the lower part of the same image, it is displayed the printed PET substrate, previously treated with $He/O_2$ plasma with a commercial plasma device. The plasma treatment increased the surface tension of the substrate of 10 Dynes/cm (passing from 38 to 48 Dynes/cm), inducing a higher wettability of the substrate.

The Fourier patterns for printed PET is of course totally different from the pattern of the substrate, because we are now observing the texture of the ink coating. The Fourier pattern estimates the grain size of texture and its anisotropy. To record OFT patterns we rotated the samples of 45 degrees. The diffractive halo of the sample with plasma-treated surface is different from that of the non-treated sample. The OFT of the plasma-treated sample has a lower anisotropy and the halo radius is smaller. This is in agreement with the higher wettability of the substrate, which allows a more homogeneous spreading of the ink on the substrate. Let us note that differences in textures between these two printed films are hardly detectable by microscopy inspection.

The same we did for the BOPP film. On the left of Fig.9, the texture of the UV ink is shown. On the right the OFT pattern. In this case too, the pattern is different from that shown in Fig.4 of BOPP substrate. In the upper part of the figure, the substrate was not subjected to a pre-treatment: the Fourier spectrum shows in its central part a trace of the substrate anisotropy. In the lower part of Fig.9, the substrate was treated with a $He/O_2$ plasma, to increase the surface tension to 58 Dynes/cm from the value of 42 Dynes/cm of the non-treated substrate. OFT is different from that previously observed. Because there are no traces of the substrate anisotropy, the ink coating is thicker.

The radius of the halo is smaller and then, as already discussed in the case of PET, this is in agreement with the higher wettability of the plasma-treated substrate. For BOPP too, differences in the textures between the two printed films are hardly detectable by microscopy inspection.

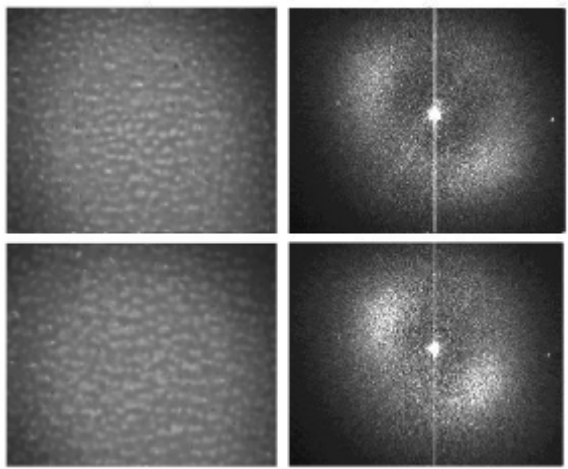

**Figue 9:** Flexo-printed BOPP films observed with microscope (image size, 1 mm) and the corresponding Fourier spectra. In the upper part of figure, the surface was not treated before printing. In the lower part, the surface was plasma treated to enhance the surface wettability. Fourier spectra are different, according to the substrate pretreatment.

## 6. CONCLUSIONS

OFTs have two potential applications in the industry of polymer films, where it can be used for on-line and off-line inspections. During the substrate production, for those polymers displaying a spectrum appreciably different from that of the light beam, OFT can monitor the structure of the film as consequence of strains induced by the processing stresses. The other application with much more potentialities, is in the detection and monitoring of ink textures.